\documentclass[12pt,preprint]{aastex}

\shorttitle{Magnetic Reconnection at Coronal Hole Boundaries}
\shortauthors{Yang, et al.}

\begin{document}

\title{ \emph{SDO} Observations of Magnetic Reconnection at Coronal
Hole Boundaries}

\author{Shuhong Yang\altaffilmark{1,3}, Jun Zhang\altaffilmark{1}, Ting Li\altaffilmark{1,3},
Yang Liu\altaffilmark{2}}

\altaffiltext{1}{Key Laboratory of Solar Activity, National
Astronomical Observatories, Chinese Academy of Sciences, Beijing
100012, China; [shuhongyang;zjun;liting]@nao.cas.cn}

\altaffiltext{2}{W.W. Hansen Experimental Physics Laboratory,
Stanford University, Stanford, CA 94305-4085, USA;
yliu@quake.stanford.edu}

\altaffiltext{3}{Graduate University of Chinese Academy of Sciences,
Beijing, China}

\begin{abstract}

With the observations from the Atmospheric Imaging Assembly (AIA)
and the Helioseismic and Magnetic Imager (HMI) aboard the
\emph{Solar Dynamics Observatory}, we investigate the coronal hole
boundaries (CHBs) of an equatorial extension of polar coronal hole.
At the CHBs, lots of extreme-ultraviolet (EUV) jets, which appear to
be the signatures of magnetic reconnection, are observed in the 193
{\AA} images, and some jets occur repetitively at the same sites.
The evolution of the jets is associated with the emergence and
cancelation of magnetic fields. We notice that both the east and the
west CHBs shift westward, and the shift velocities are close to the
velocities of rigid rotation compared with those of the photospheric
differential rotation. This indicates that magnetic reconnection at
CHBs results in the evolution of CHBs and maintains the rigid
rotation of coronal holes.

\end{abstract}


\section{Introduction}

Coronal holes (CHs) are dark areas which can be observed at both the
low latitudes and the polar regions of the Sun using X-ray and
extreme-ultraviolet (EUV) lines (Chiuderi Drago et al. 1999). The
magnetic fields within a CH region are dominated by one polarity,
and thus the field lines in the upper atmosphere are open to the
interplanetary region, generating high-speed solar winds that can
lead to geomagnetic storms (Bohlin 1977; Krieger \& Timothy 1973; Tu
et al. 2005). Many basic characters of CHs and their relationship
with magnetic fields have been studied (Wiegelmann \& Solanki 2004;
Meunier 2005; Zhang et al. 2006, 2007; Yang et al. 2009a, b). Using
observations from the Solar Ultraviolet Measurements of Emitted
Radiation (SUMER) instrument onboard the \emph{Solar and
Heliospheric Observatory} (\emph{SOHO}), blue shifts indicating
plasma flows within CHs are intensively investigated (Hassler et al.
1999; Peter \& Judge 1999; Xia et al. 2004; Aiouaz et al. 2005;
McIntosh et al. 2011). Recently, \emph{Hinode} (Kosugi et al. 2007)
observations display that CHs are nonpotential, e.g., in some strong
magnetic field regions within CHs, the current densities are as
large as those in flare productive active regions (Yang et al.
2011).

CHs can be classified into two types according to their locations:
polar CHs and mid-latitude ones (Insley et al. 1995; Wang et al.
1996). The mid-latitude CHs can be ``isolated" or connected with
polar CHs called equatorial extensions of polar CHs (EECHs) (Insley
et al. 1995). Observations reveal that EECHs rotate quasi-rigidly
although the photosphere rotates differentially (Timothy et al.
1975; Insley et al. 1995; Wang et al. 1996). CH boundaries (CHBs)
separate two kinds of areas with different configurations: CHs with
open magnetic fields and surrounding quiet Sun with closed magnetic
fields. Due to the fact that EECHs rotate rigidly while the
underlying photospheric fields differentially, magnetic reconnection
is necessary to happen at CHBs to maintain the rigid rotation of the
CHs (Wang \& Sheeley 1994; Fisk, et al. 1999).

Kahler \& Moses (1990) studied an EECH in \emph{Skylab} X-ray images
and noticed that X-ray bright points at CHBs play an important role
in the CH expansion and contraction. Kahler \& Hudson (2002) studied
three EECHs observed by the \emph{Yohkoh} Soft X-ray telescope,
however, they found no significant effect for bright points on CHB
evolution. Using EUV images from the \emph{Transition Region and
Coronal Explorer} (\emph{TRACE}) and the \emph{SOHO}, Madjarska \&
Wiegelmann (2009) investigated the evolution of CHBs at small
scales. They found that small-scale magnetic loops (appeared as
bright points) play an important role in CHB evolution but did not
find signature for a major energy release during the evolution of
loops. Kahler et al. (2010) also focused on the changes of CHBs. But
in their study, no example of the energetic jet was reported in EUV
images. According to previous studies, jets are transient plasma
ejections which are believed to result from magnetic reconnection
(Shibata et al. 1992, 1994; Yakoyama \& Shibata 1995). More
recently, Subramanian et al. (2010) explored CHB evolution with the
images from XRT on board \emph{Hinode}. They observed some jet-like
events at CHB regions and those ejections appeared to be triggered
by magnetic reconnection.

The aim of this study is to examine if there exist some signatures
(e.g., jets) of magnetic reconnection at CHBs in the EUV images from
the \emph{Solar Dynamics Observatory} (\emph{SDO}; Schwer et al.
2002). In Sections 2 and 3, the observations and results are
presented respectively. The conclusions and discussion are given in
Section 4.

\section{Observations and Data Analysis}

The CH studied here is an EECH extended from the south pole to
N41$\degr$ in the middle of 2010 June (see Figure 1\emph{b}). The
data used here were obtained by the Atmospheric Imaging Assembly
(AIA; Title et al. 2006) and the Helioseismic and Magnetic Imager
(HMI; Schou et al. 2011) aboard the \emph{SDO} from June 10 to June
15. The AIA takes full-disk images in 10 wavelengths with pixel size
of 0$\arcsec$.6 and we use the 193 {\AA} data at Level 1.5 with 12 s
cadence to study the CH. We also adopt the full-disk line-of-sight
magnetograms with 45 s cadence obtained by the HMI at 6173 {\AA}
with a spatial sampling of 0$\arcsec$.5 pixel$^{-1}$. The 193
channel contains lines from both Fe XII and Fe XXIV, and the
corresponding temperatures are 1.5 MK and 20 MK, respectively. Among
10 wavelengths of AIA, 193 {\AA} mainly reveals information of
corona, 304 {\AA} and 171 {\AA} are mainly formed in chromosphere
and transition region, while other lines are sensitive in response
to the active region or flaring corona. Since CHs are low density
and low temperature regions in the corona, so we choose 193 {\AA} as
the most appropriate one to study CHs. We can find that CHs are
clearly visible in 193 {\AA} images.

The AIA images and HMI magnetograms are differentially rotated to a
reference time (2010 June 13 00:00:06 UT) when the CH was mainly
located at the central meridian. In this study, the AIA images are
displayed on logarithmic scale and the HMI magnetograms are averaged
every three frames. The CHBs are determined as the regions with
intensities 1.5 times the average intensity of the darkest region
inside the CH, the same as the method that Madjarska \& Wiegelmann
(2009) used. We define the CHB regions as $\pm$ 15$\arcsec$ on both
sides of the CHBs, as Subramanian et al. (2010) did in their study.

\section{Results}

Some parts of the CHB are diffuse, so we select two parts (Figures
1\emph{a} and 1\emph{c}, also outlined by two squares in Figure
1\emph{b}) which are sharp enough to be defined well. In the early
morning of 2010 June 13, the main body of the CH was mainly at the
central meridian. To avoid projection effects (Kahler \& Hudson
2002; Kahler et al. 2010), we use only the limbward CHB
observations, i.e., the east CHB (ECHB) on June 12 (Figure
1\emph{a}) and the west CHB (WCHB) on June 13 (Figure 1\emph{c}).

\subsection{Signatures of Magnetic Reconnection at the CHBs}

The background of Figure 1\emph{a} is an AIA 193 {\AA} image at
00:00:06 UT on 2010 June 13. The sub-regions outlined by white
rectangles display the jets occurring at different times on June 12,
and the time of each jet is also listed in the Figure. In a one-day
period, seven sites with bright jets are observed at the ECHB, as
presented in Figure 1\emph{a}. At the WCHB, more jets appeared
(Figure 1\emph{c}). On June 13, we find fifteen sites at the WCHB
where obvious jets occurred. Here, we identify the jets by eye
according to their appearance and evolution in 193 {\AA} movies,
instead of with automatic techniques. The automatic detection
methods introduced in the newly published paper of Martens et al.
(2011) will be helpful for us to investigate the jets at CHBs in
future.

Some jets occurred repetitively at the same sites (e.g., in
rectangle ``8" in Figure 1\emph{c}). Along slit ``A---B", we obtain
the image profile by averaging five pixels in the running ratio
images in the direction perpendicular to ``A---B". Then we make the
running ratio space-time plot of such profiles over time from 17:00
UT to 21:00 UT, as shown in Figure 1\emph{d}. During this period,
three obvious jets are observed, as labeled with ``i", ``ii", and
``iii," and their lifetimes are about 12 minutes. The projection
velocities of these three jets are 61 km s$^{-1}$, 61 km s$^{-1}$,
and 54 km s$^{-1}$, respectively.

\subsection{Evolution of the Jets at the CHBs}

The sequence of 193 {\AA} images in Figure 2 (from \emph{a} to
\emph{d}) display the evolution of two jets located at the WCHB
(also see rectangle ``5" in Figure 1\emph{c}). The underlying
magnetic fields are presented in Figures 2\emph{e}--\emph{h}. At
16:44:03 UT, a cluster of magnetic elements with mixed polarities
(denoted by arrow ``3") emerged at the CHB, and the negative flux,
opposite to the surrounding polarities, became larger at 18:44:03
UT. During this process, bright point in the 193 {\AA} images was
observed at the location of magnetic emergence (indicated by arrow
``1"). The newly emerged negative elements moved toward the
surrounding positive ones and canceled with them (panels
\emph{g}--\emph{h}). Above the cancelation region, two jets (denoted
by arrows ``2") occurred in the corona (panels \emph{b}--\emph{c})
and then decayed (panel \emph{d}).

Figure 3 presents the evolution process of another jet. At 20:09:06
UT, the area outlined by the square was a CHB region without obvious
coronal activity (Figure 3\emph{a}). Several minutes later, a jet
appeared (denoted by arrow ``1"). At 20:20:18, the jet had become
more violent (Figure 2\emph{c}). The decay of the jet is displayed
in Figures 3\emph{d}--\emph{e}, and at 20:59:18 UT (Figure
3\emph{f}), the jet disappeared. Correspondly, there are some
changes in the underlying magnetic fields (Figures
3\emph{g}--\emph{i}). At 20:08:03 UT, before the occurrence of the
jet, there existed lots of magnetic elements with opposite
polarities (denoted by arrows labeled ``2"). During the jet
evolution, the negative elements canceled with the positive ones
(Figure 3\emph{h}). At 21:29:03 UT, the total magnetic flux
decreased obviously and only several minor elements were remained
(Figure 3\emph{i}).

\subsection{Shifts of the CHBs }

In order to analyze the shifts of the CHBs, the derotated sub-images
outlined by two dashed squares in Figure 1 are mapped into
heliographic coordinates and used in Figures 4 and 5.

The FOV of Figure 4 covers a part of sharp ECHB. Panels
\emph{a}--\emph{d} show the changes of ECHB on June 12. We can find
some jets occurred at the ECHB (denoted by arrows ``1" and ``2"; see
also Figure 1\emph{a}). To examine the shift of ECHB, we make three
space-time plots along three slits (``A---B", ``C---D", and ``E---F"
in Figure 4\emph{b}) in the latitudinal direction, and display them
in Figures 4\emph{e}--\emph{g}, respectively. In the plots, each
profile is obtained by averaging five pixels in 193 {\AA} images in
the direction perpendicular to the slits. The shift velocity of the
ECHB at each site is derived by a linear fitting. We find that the
shift velocities at three sites are 0$\degr$.755 $\pm$ 0$\degr$.008
day$^{-1}$, 0$\degr$.619 $\pm$ 0$\degr$.004 day$^{-1}$, and
0$\degr$.369 $\pm$ 0$\degr$.003 day$^{-1}$, respectively. The
feature denoted by arrow ``3" in panel \emph{f} represents a jet.
However, we find no obvious jet in panels \emph{e} and \emph{g}.

We also study a sharp section of WCHB, which is presented in Figure
5. Figures 5\emph{a}--\emph{d} exhibit the WCHB at different stages
on June 13. Arrows ``1" and ``2" denote two jets appearing at the
WCHB and more jets can be found in Figure 1\emph{c}. Similarly as we
deal with the ECHB, space-time plots along the slits ``A---B",
``C---D", ``E---F" (see Figure 5\emph{b}) are obtained and given in
Figures 5\emph{e}--\emph{g}. In the three plots, we obtain the
fitted velocities: 0$\degr$.789 $\pm$ 0$\degr$.009 day$^{-1}$,
1$\degr$.015 $\pm$ 0$\degr$.005 day$^{-1}$, and 0$\degr$.949 $\pm$
0$\degr$.007 day$^{-1}$, respectively. The bright features indicated
by arrows ``3"--``5" are jets at the WCHB.

\section{Conclusions and Discussion}

By employing the observations from the AIA and HMI on board the
\emph{SDO}, we investigate the CHBs of an EECH. At the CHBs, a lot
of jets are found in the 193 {\AA} images, and some jets occurred
repetitively at the same sites. The evolution of the jets is
associated with emergence and cancelation of the underlying magnetic
fields. We notice that both the ECHB and the WCHB shift westward,
and the shift velocities are close to the velocities of rigid
rotation compared with those of the photospheric rotation.

Subramanian et al. (2010) observed X-ray jets at CHBs
with \emph{Hinode} XRT data and they thought that those ejections
appeared to be triggered by magnetic reconnection. In this study, we
have observed many bright jets at both the ECHB and the WCHB in the
EUV line images from the \emph{SDO}. We think that these EUV jets
are the observational signatures of magnetic reconnection at CHBs.
In this study, more jets are observed at the WCHB than at the ECHB.
It seems that this phenomenon is consistent with the concept of the
E---W asymmetry in the CHB reconnection as discussed in Kahler et
al. (2010).

We notice that some jets occurred repetitively at the same sites
(see Figure 1\emph{d}). The recurrence of the jets at CHBs is
similar to the events in CHs, quiet regions, and active regions.
Wang et al. (1997) and Chae et al. (1998, 1999) studied the disk and
limb jets in quiet and active regions and found that the jets
occurred repeatedly at the same sites. Repetitive occurrence of jets
in CHs is also presented in the study of Subramanian et al. (2010).
They reported that many bright points produced several jet-like
events with no periodicity.

The average projection velocity of the jets in this study is about
60 km s$^{-1}$. It is difficult to know the real velocities of the
jets from the projection velocities in this study. But if we assume
the jets are locally vertical and combine with their heliocentric
angle of almost 30$\degr$, the real velocities should be around 120
km s$^{-1}$, about 2 times the projection velocities. Cirtain et al.
(2007) studied four jets in detail and found there are two velocity
components for each jet. one component is slow velocity of $\sim$200
km s$^{-1}$, close to the sound speed, and the other one observed at
the start of each event is roughly $\sim$800 km s$^{-1}$, close to
the Alfv\'{e}n speed. They considered them as two kinds of outflows
during the post-magnetic reconnection phase of a jet. Savcheva et
al. (2007) statistically studied a large sample of jets and found
that the outward velocity shows a peaked distribution with a maximum
at 160 km s$^{-1}$. The estimated real velocity (120 km s$^{-1}$) in
our study is generally consistent with those of Cirtain et al.
(2007; 200 km s$^{-1}$) and many other former studies (Shibata et
al. 1992; Shimojo et al. 1996; Savcheva et al. 2007). So we think
that the jets with low velocities in our study are also resulted
from magnetic reconnection, since they have been considered as one
kind of outflows during the post-magnetic reconnection phase
(Cirtain et al. 2007), though they might be directly driven by
magnetosonic shock waves. In this study, we do not derive high
velocity of about 800 km s$^{-1}$. Savcheva et al. (2007) introduced
a ``brightness contour method" which can detect high velocities of
about 600--1000 km s$^{-1}$, but the error bars on the high
velocities are usually very large. However, this method can be
considered to use to study jets in detail in future.

In the studies of Madjarska \& Wiegelmann (2009) and Subramanian et
al. (2010), no photospheric magnetogram is used. So the evolution of
magnetic fields associated with the bright points and jets at CHBs
is not clear. With the HMI observations, we are able to know the
evolution process of the magnetic fields underlying the coronal jets
at the CHBs. When the jets occurred, magnetic emergence and
cancelation were observed, as shown in Figures 2 and 3.

Since the AIA images are differentially rotated to a reference time,
the CHBs should shift westward if the CH rotates rigidly and should
keep stable without shift if the CH rotates differentially with the
photosphere. After calculation, we know that, if the CH rotates
rigidly, the shift velocity of the ECHB shown in Figure 4 should be
0$\degr$.669 day$^{-1}$ and that of the WCHB in Figure 5 should be
0$\degr$.490 day$^{-1}$. Actually, we obtain that the mean
velocities of the ECHB and the WCHB are 0$\degr$.581 day$^{-1}$ and
0$\degr$.918 day$^{-1}$, respectively. The observed shift velocities
are close to those required to maintain the rigid rotation compared
with those of the photospheric differential rotation. It provides us
exact evidences of the concept that magnetic reconnection at CHBs
results in the evolution of CHBs and maintains the rigid rotation of
CHs. The shift velocity of the WCHB is higher than the rigid
rotation requires, which may be caused by the intensive magnetic
reconnection indicated by more jets at the WCHB during this period.

However, not all the CHB shifts are accompanied by obvious
signatures (such as jets) of magnetic reconnection, as revealed in
Figures 4\emph{e} and \emph{g}. We think this may be caused by the
concept that the reconnection at CHBs occurs just below the
potential-field source surface (PFSS) (Wang \& Sheeley 1994; Fisk
2005; Schwadron et al. 2005; Wang et al. 2007). This explanation is
also adopted in the study of Kahler et al. (2010). They argued that
gradual reconnection occurs at high altitudes, so they did not find
jets and flare-like events in the EUV images. In our opinion, short
period (about seven hours) and low cadence (4.4 min) observations
could lead to the absence of jets in the EUV images. In addition,
another possible reason is that magnetic reconnection at some sites
is too weak to be observed. Based on the results, we think magnetic
reconnection at CHBs can take place in the low corona indicated by
the jets in 193 {\AA} images, also in the high layer just below the
PFSS as mentioned above.

In this Letter, we present the general appearance of the signatures,
i.e. the jets in 193 {\AA} images, of magnetic reconnection at the
CHBs. But there are still lots of questions. Does the reconnection
occur between open field lines in the CH and closed loops in the
surrounding quiet Sun regions? Where does the reconnection take
place and in which layer are the jets accelerated? Is it in corona
or the lower regions? Is it possible that the formation of jets is
caused by reconnection between low lying loops in the CH which can
not extend in to the corona and significantly high coronal loops in
the quiet Sun? As pointed out by the referee, observations in the
other cooler wavelengths formed in chromosphere and transition
region might help to get more insights regarding reconnection at
CHBs, which can be taken into consideration in our next study.

\acknowledgments {We thank the referee for the helpful comments.
\emph{SDO} is a mission for NASA's Living With a Star (LWS) Program.
This work is supported by the National Natural Science Foundations
of China (40890161, 11025315, 10921303, 41074123 and 11003024), the
CAS Project KJCX2-YW-T04, the National Basic Research Program of
China under grant 2011CB811403, and the Young Researcher Grant of
National Astronomical Observatories, Chinese Academy of Sciences.}

{}

\clearpage

\begin{figure}
\centering
\includegraphics
[bb=23 219 580 509,clip,angle=0,scale=0.85]{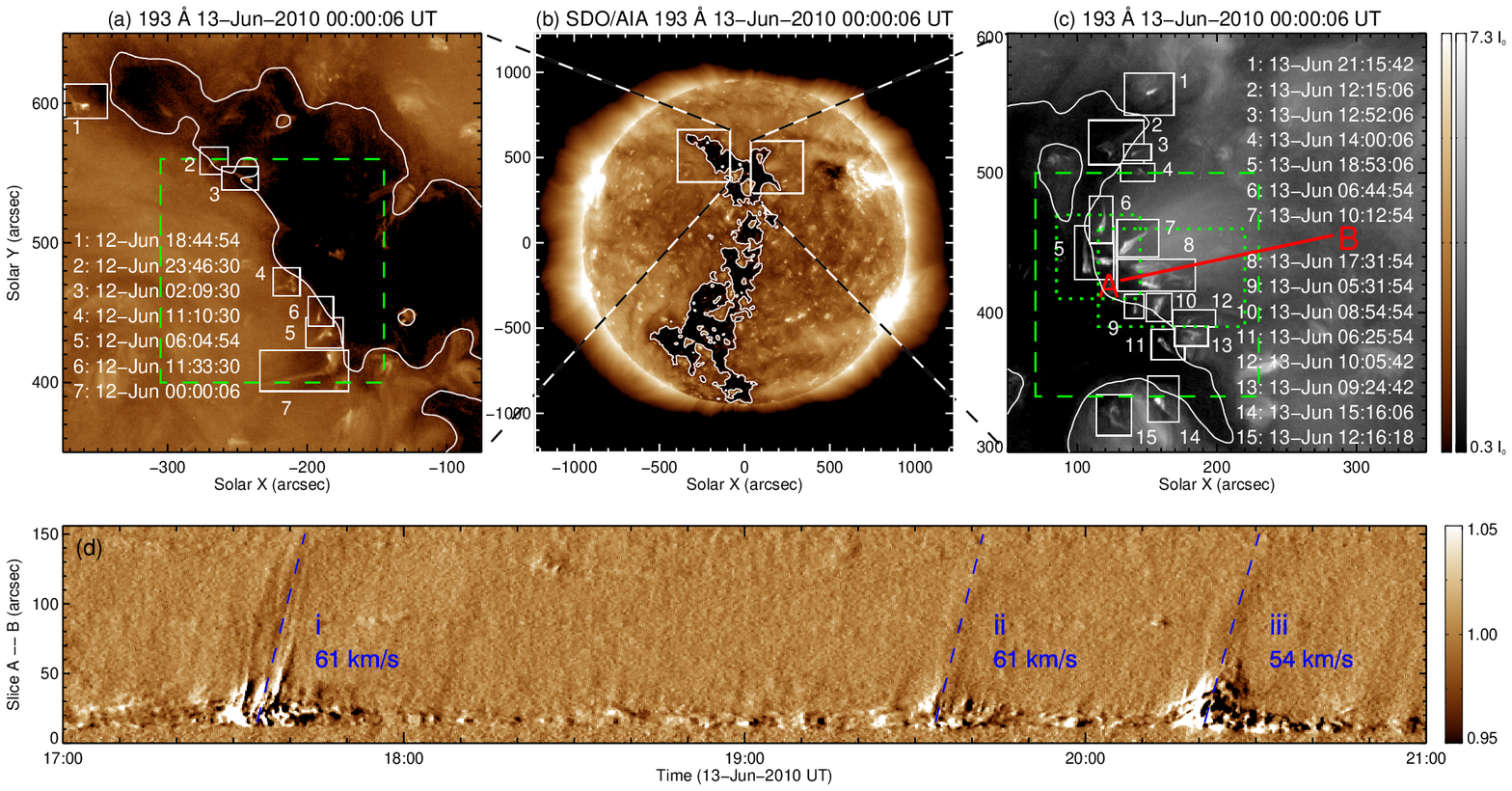} \caption{
{\emph{SDO}}$/$AIA 193 {\AA} full-disk image (panel \emph{b})
obtained on 13 June 2010 showing an EECH, and sub-images (panels
\emph{a} and \emph{c}) with sharp CHBs. Panel \emph{d} is running
ratio space-time plot along slit ``A---B" (red line in panel
\emph{c}). The contours represent CHBs. The white rectangles in
panels \emph{a} and \emph{c} show jets occurring at the CHBs at
different times. The dotted rectangles (left and right) in panel
\emph{c} outline the FOVs of Figures 2 and 3, respectively, and two
dashed squares in panels \emph{a} and \emph{c} the FOVs of Figures 4
and 5, respectively. \label{fig1}}
\end{figure}
\clearpage

\begin{figure}
\centering
\includegraphics
[bb=110 280 506 470,clip,angle=0,scale=1.1]{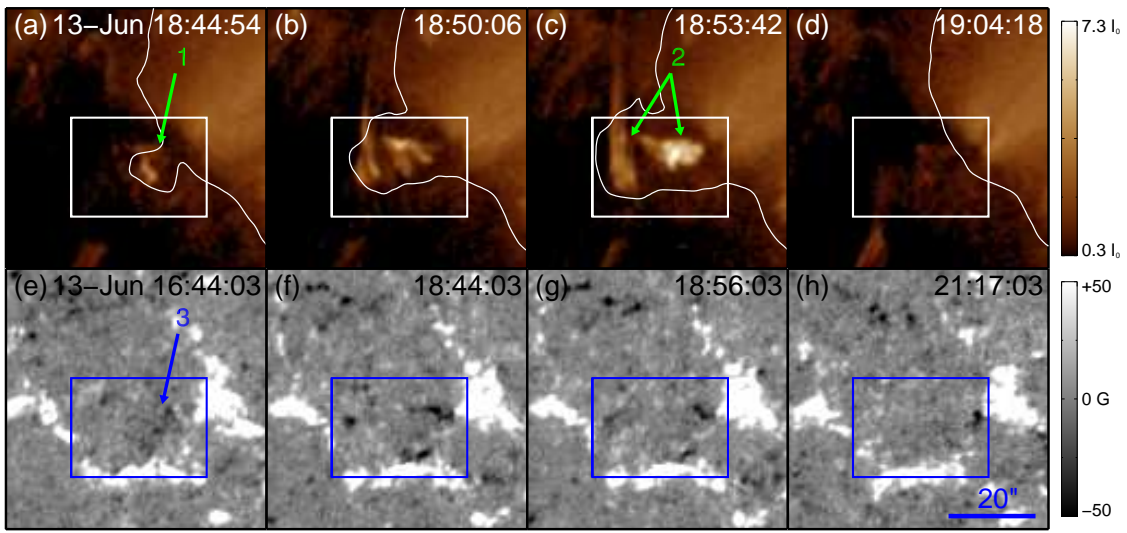}
\caption{Sequence of AIA 193 {\AA} images (upper panels) displaying
the evolution of two jets, and HMI longitudinal magnetograms (lower
panels) showing the underlying magnetic fields. The curves delineate
CHBs and the rectangles outline the area within which magnetic
emergence and cancelation took place. Arrows ``1", ``2", and ``3"
denote the bright point, jets, and emerging magnetic flux,
respectively. \label{fig2}}
\end{figure}
\clearpage

\begin{figure}
\centering
\includegraphics
[bb=154 290 462 487,clip,angle=0,scale=1.2]{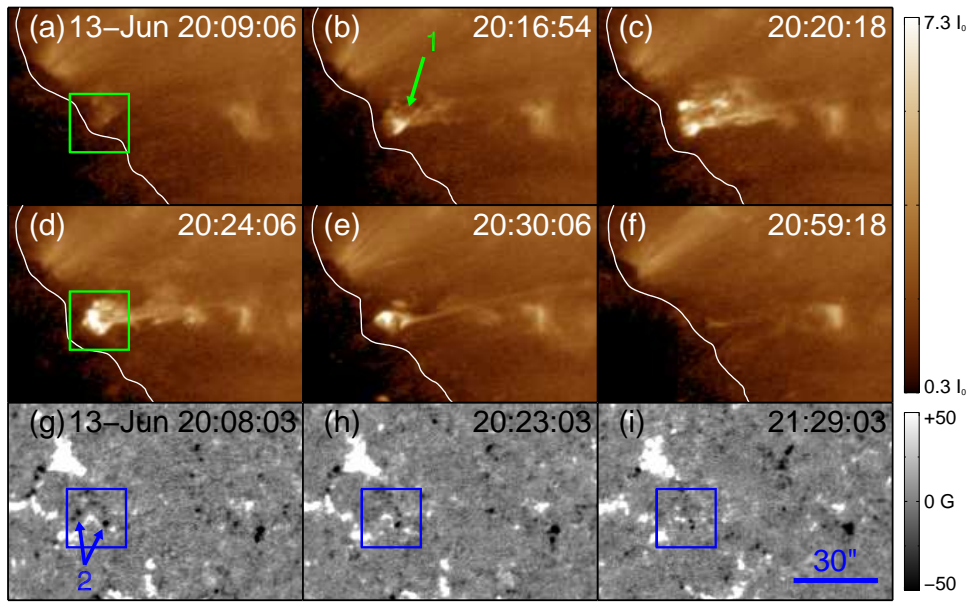} \caption{AIA
images (panels \emph{a}--\emph{f}) and HMI magnetograms (panels
\emph{g}--\emph{i}) exhibiting a signature of magnetic reconnection.
The curves delineate CHBs and the squares outline the area where
magnetic cancelation occurred. Arrows ``1" and ``2" indicate the jet
and negative magnetic elements, respectively. \label{fig3}}
\end{figure}
\clearpage

\begin{figure}
\centering
\includegraphics
[bb=42 200 555 500,clip,angle=0,scale=0.92]{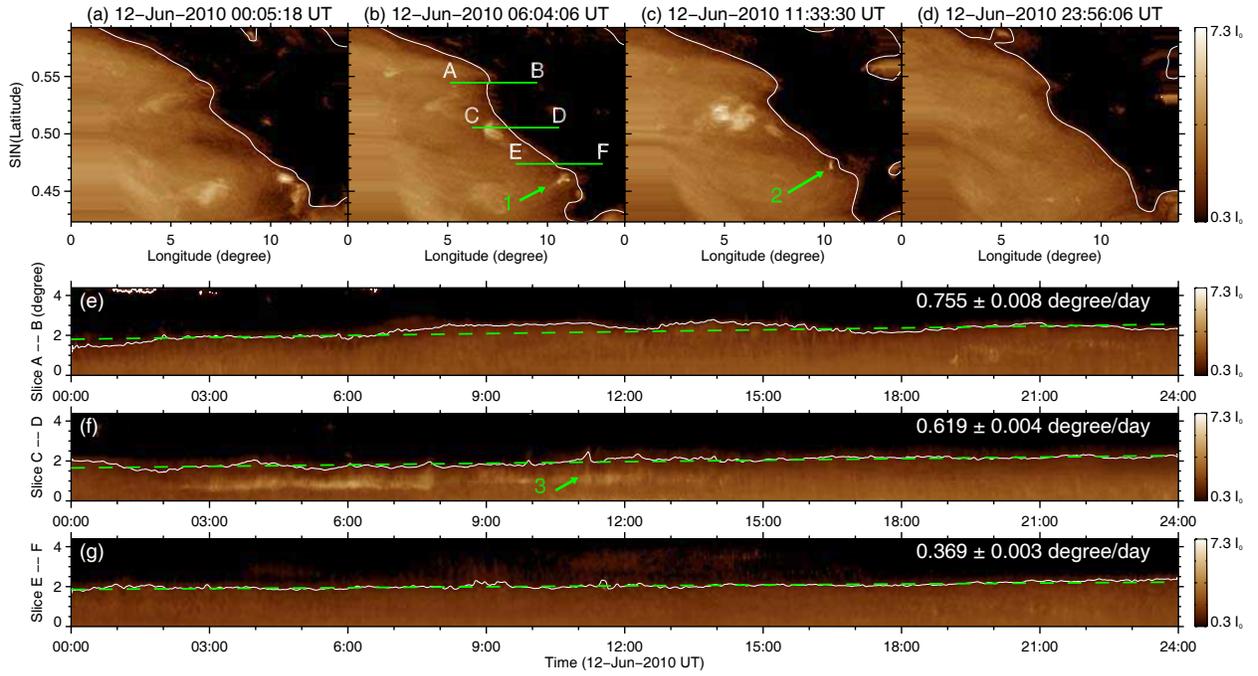} \caption{ECHB
shift revealed in 193 {\AA} images mapped into heliographic
coordinates (panels \emph{a}--\emph{d}) and space-time plots (panels
\emph{e}--\emph{g}) along slits ``A---B", ``C---D", and ``E---F"
(see panel \emph{b}), respectively. The curves delineate CHBs and
the arrows denote jets. The dashed lines are linear fits to the
CHBs. \label{fig4}}
\end{figure}
\clearpage

\begin{figure}
\centering
\includegraphics
[bb=42 170 555 500,clip,angle=0,scale=0.92]{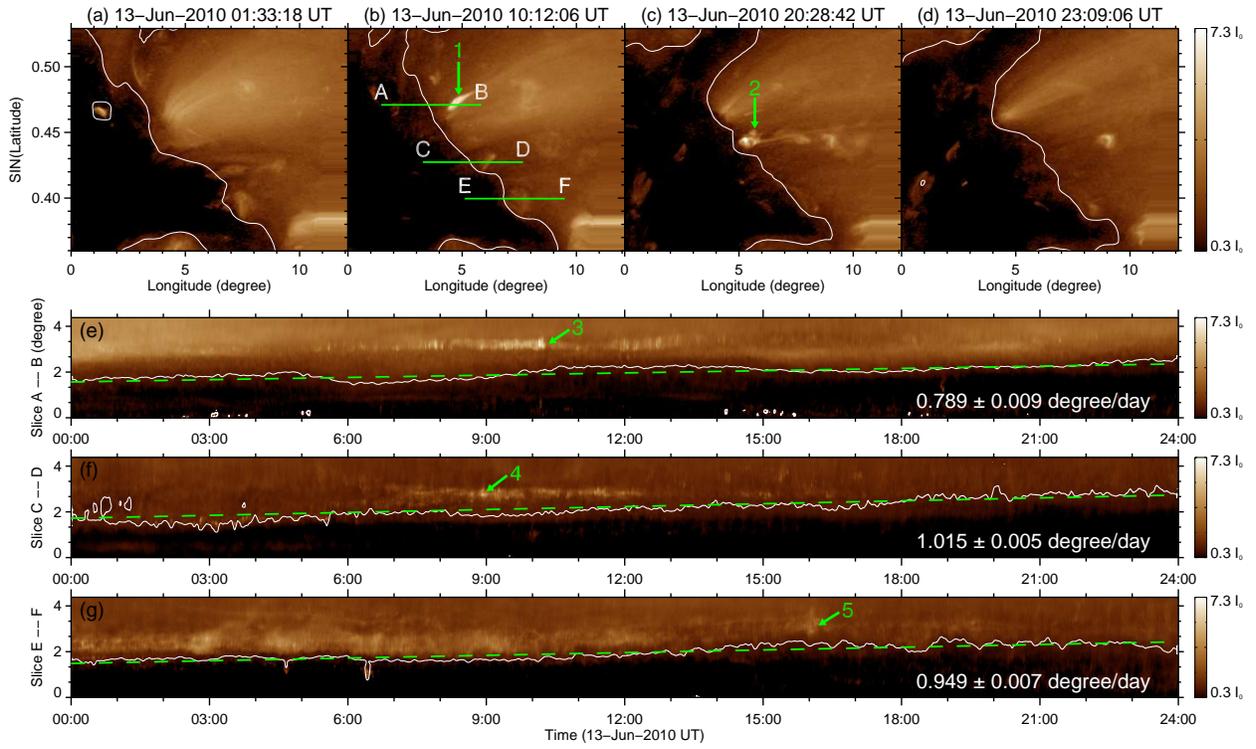} \caption{Similar
to Figure 4 but for the WCHB. \label{fig5}}
\end{figure}
\clearpage

\end{document}